\documentclass[12pt]{article}
\usepackage{graphicx}
\newcommand{\be}{\begin{equation}}
\newcommand{\ee}{\end{equation}}
\newcommand{\bea}{\begin{eqnarray}}
\newcommand{\eea}{\end{eqnarray}}

\newcommand{\xbj}{x_{\!\scriptscriptstyle B}}

\newcommand{\bfk}{\mbox{\boldmath $k$}}
\newcommand{\bfq}{\mbox{\boldmath $q$}}
\newcommand{\bfP}{\mbox{\boldmath $P$}}

\newcommand{\pup}{p^\uparrow}

\newcommand{\bfp}{\mbox{\boldmath $p$}}
\newcommand{\bfz}{\mbox{\boldmath $z$}}
\newcommand{\bfS}{\mbox{\boldmath $S$}}

\newcommand{\nd}{\noindent}

\def\lsim{\mathrel{\rlap{\lower4pt\hbox{\hskip1pt$\sim$}}\raise1pt\hbox{$<$}}}
\def\gsim{\mathrel{\rlap{\lower4pt\hbox{\hskip1pt$\sim$}}\raise1pt\hbox{$>$}}}
\def\nostrocostruttino#1\over#2{\mathrel{\mathop{\kern 0pt \rlap
{\hbox{$#1$}}} \hbox{\kern-.135em $#2$}}}

\newcommand{\NP}[1]{{\it Nucl.\ Phys.}\ {\bf #1}}

\newcommand{\PL}[1]{{\it Phys.\ Lett.}\ {\bf #1}}
\newcommand{\PR}[1]{{\it Phys.\ Rev.}\ {\bf #1}}
\newcommand{\PRL}[1]{{\it Phys.\ Rev.\ Lett.}\ {\bf #1}}

\def\kt{k_\perp}

\def\ptq{p_\perp}

\begin{document}
\begin{flushright} 
\end{flushright} 
\vskip 1.5cm
\begin{center}
{\bf Extracting the Sivers function from polarized SIDIS data\\
and making predictions} 
\\
\vskip 1.2cm
{\sf M.~Anselmino$^1$, M.~Boglione$^1$, U.~D'Alesio$^2$, A. Kotzinian$^3$, 
F.~Murgia$^2$, A. Prokudin$^1$}
\vskip 0.5cm
{\it $^1$Dipartimento di Fisica Teorica, Universit\`a di Torino and \\
          INFN, Sezione di Torino, Via P. Giuria 1, I-10125 Torino, Italy}\\
\vspace{0.3cm}
{\it $^2$INFN, Sezione di Cagliari and Dipartimento di Fisica,  
Universit\`a di Cagliari,\\
C.P. 170, I-09042 Monserrato (CA), Italy}\\ 
\vspace{0.3cm}
{\it $^3$Dipartimento di Fisica Generale, Universit\`a di Torino and \\
          INFN, Sezione di Torino, Via P. Giuria 1, I-10125 Torino, Italy}
\end{center}

\vspace{1.5cm}

\begin{abstract}
\noindent
The most recent data on the weighted transverse single spin asymmetry 
$A_{UT}^{\sin(\phi_h-\phi_S)}$ from HERMES and COMPASS 
collaborations are analysed within LO parton model with unintegrated parton
distribution and fragmentation functions; all transverse motions are taken 
into account, with exact kinematics, in the elementary interactions. 
The overall quality of the data is such that, for the first time, a rather 
well constrained extraction of the Sivers function for $u$ and $d$ quarks is 
possible and is performed. Comparisons with models are made. Based on the 
extracted Sivers functions, predictions for $A_{UT}^{\sin(\phi_h-\phi_S)}$ 
asymmetries at JLab are given; suggestions for further measurements at 
COMPASS, with a transversely polarized hydrogen target and selecting 
favourable kinematical ranges, are discussed. Predictions are also presented 
for Single Spin Asymmetries (SSA) in Drell-Yan processes at RHIC and GSI.  
\end{abstract}
\vspace{0.6cm}

\newpage 
\pagestyle{plain} 
\setcounter{page}{1} 
\nd 
{\bf 1. Introduction } 
\vskip 6pt 
In a recent paper \cite{ourpaper} we have discussed the role of intrinsic 
motions in inclusive and Semi-Inclusive Deep Inelastic Scattering (SIDIS) 
processes, both in unpolarized and polarized $\ell \, p \to \ell \, h \, X$ 
reactions. The LO QCD parton model computations have been compared with the 
experimental dependence of the unpolarized cross section on the azimuthal 
angle, around the virtual photon direction, between the leptonic and the 
hadronic planes (Cahn effect \cite{cahn}); at small transverse momentum 
$P_T$ of the produced hadron $h$, such an effect is dominantly related to 
intrinsic motions and it allows an estimate of the average values of the 
transverse momenta of quarks inside a proton, $\bfk_\perp$, and of final 
hadrons inside the fragmenting quark jet, $\bfp_\perp$, with the best fit 
results:
\be
\langle\kt^2\rangle   = 0.25  \;({\rm GeV}/c)^2 \quad\quad\quad 
\langle\ptq^2\rangle  = 0.20 \;({\rm GeV}/c)^2 \>. 
\label{ktpar}
\ee

More detail, both about the kinematical configurations and conventions 
\cite{trento} and the fitting procedure can be found in Ref. \cite{ourpaper}. 
We only notice here that the above values have been derived from sets of data 
collected at different energies and in different ranges of the kinematical 
variables $\xbj$, $Q^2$ and $z_h$, looking at the combined production of all 
charged hadrons in SIDIS processes; constant and flavour independent values 
of $\langle\kt^2\rangle$ and $\langle\ptq^2\rangle$ have been assumed, 
avoiding at this stage complications related to possible $x, z$ and $Q^2$ 
dependences. Rather than a definite derivation, the above results are better 
considered as a consistent simple estimate and a convenient parameterization 
of the true intrinsic motion of quarks in nucleons and of hadrons in jets, 
supported by the available experimental information.

Equipped with such estimates, in Ref. \cite{ourpaper} we have studied the 
transverse single spin asymmetries $A_{UT}^{\sin(\phi_{\pi}-\phi_S)}$ observed 
by HERMES collaboration \cite{hermUT}; that allowed a first rough extraction 
of the Sivers function \cite{siv}
\be 
\Delta^N \! f_ {q/\pup}(x,k_\perp) = - \frac{2\,k_\perp}{m_p} \> 
f_{1T}^{\perp q}(x, k_\perp) \>, \label{rel} 
\ee
defined by
\bea
f_ {q/\pup} (x,\bfk_\perp) &=& f_ {q/p} (x,\kt) +
\frac{1}{2} \, \Delta^N \! f_ {q/\pup}(x,\kt)  \;
{\bfS} \cdot (\hat {\bfP}  \times
\hat{\bfk}_\perp) \label{sivnoi} \\
&=& f_ {q/p} (x,\kt) - f_{1T}^{\perp q} (x,\kt)  \;
\frac {{\bfS} \cdot (\hat {\bfP}  \times
{\bfk}_\perp)}{m_p} \> , \label{sivpiet}
\eea
where $f_ {q/p}(x,\kt)$ is the unpolarized $x$ and $\kt$ dependent 
parton distribution ($\kt = |\bfk_\perp|$); $m_p$, $\bfP$ and $\bfS$ are 
respectively the proton mass, momentum and transverse polarization vectors
($\hat{\bfP}$ and $\hat{\bfk}_\perp$ denote unit vectors). 

The Sivers function extracted from HERMES data in \cite{ourpaper} was shown 
to be consistent with preliminary COMPASS data on 
$A_{UT}^{\sin(\phi_h-\phi_S)}$ obtained on a deuteron target, in a different 
kinematical region \cite{compUT}.

While the preliminary HERMES data \cite{hermUT} offered a definite indication 
of a non zero Sivers effect, their amount and quality were not yet such that 
an accurate extraction of the Sivers functions was possible; that reflects 
in the values of the parameters of the Sivers functions given in Table I of 
Ref. \cite{ourpaper}, which have large uncertainties. 

New HERMES data are now available \cite{hermnew}; they are consistent with 
the previous ones, with much smaller errors. Similarly, COMPASS collaboration 
published \cite{compnew} their preliminary results. We consider here these 
whole new sets of HERMES and COMPASS data and perform a novel fit of the 
Sivers functions. It turns out that the new data constrain much better the 
parameters, thus offering the first direct significant estimate of the 
Sivers functions -- for $u$ and $d$ quarks -- active in SIDIS processes.
The sea quark contributions are found to be negligible, at least in the 
kinematical region of the available data.  

The modeled and extracted Sivers functions $\Delta^N\!f_ {u/\pup}(x,k_\perp)$
and $\Delta^N\!f_ {d/\pup}(x,k_\perp)$ are used to compute, and thus predict, 
the values of $A_{UT}^{\sin(\phi_h-\phi_S)}$ expected at COMPASS, for 
scattering off a polarized hydrogen (rather than deuteron) target, which 
avoids cancellations between the opposite $u$ and $d$ contributions. 
Suggestions for the selection of favourable kinematic regions, where the 
asymmetry is sizeable, are discussed. Similar predictions, with strongly 
encouraging results, are given for polarized SIDIS processes at JLab. 

Finally, we exploit the QCD prediction \cite{col} 
\be
f_{1T}^{\perp q}(x, k_\perp)|_{\rm{DIS}} = -
f_{1T}^{\perp q}(x, k_\perp)|_{\rm{D-Y}} \label{DIS-DY} 
\ee
and compute a single spin asymmetry, which can only originate from the Sivers
mechanism \cite{noi2}, for Drell-Yan processes at RHIC and GSI.      

\vspace{18pt}
\goodbreak
\nd
{\bf 2. Extracting the Sivers functions}
\vskip 6pt 
Following Ref. \cite{ourpaper}, the inclusive ($\ell \, p \to \ell \, X $)
unpolarized DIS cross section in non collinear LO parton model is given by 
\be
\frac{d^2 \sigma ^{\ell p\to \ell X}}{d \xbj \, dQ^2} = 
\sum_q \int {d^2 \bfk _\perp}\; f_q(x,\kt) \; 
\frac{d\hat\sigma ^{\ell q\to \ell q}}{dQ^2} \;
J(\xbj, Q^2, \kt) \>,
\label{dsigkt}
\ee
and the semi-inclusive one ($\ell \, p \to \ell \, h \, X $) by
\be
\frac{d^5\sigma^{\ell p \to \ell h X }}{d\xbj \, dQ^2 \, dz_h \, d^2 \bfP _T} =
\sum_q \int {d^2 \bfk _\perp}\; f_q(x,\kt) \;
\frac{d \hat\sigma ^{\ell q\to \ell q}}{dQ^2} \;
J\; \frac{z}{z_h} \; D_q^h(z,p _\perp) \>,
\label{sidis-Xsec-final} 
\ee 
where  
\be
J = \frac{\hat s^2}{\xbj^2 s^2} \; \frac{\xbj}{x}
\left( 1 + \frac{\xbj^2}{x^2}\frac{\kt^2}{Q^2} \right)^{\!\!-1}  
\label{fcnj}
\ee
and
\be
\frac{d \hat\sigma^{\ell q\to \ell q}}{d Q^2} = e_q^2 \, 
\frac{2\pi \alpha^2}{\hat s^2}\,
\frac{\hat s^2+\hat u^2}{Q^4}\;\cdot
\label{part-Xsec}
\ee

$Q^2$, $\xbj$ and $y = Q^2/(\xbj\,s)$ are the usual leptonic DIS variables 
and $z_h, \bfP_T$ the usual hadronic SIDIS ones, in the $\gamma^*$--$p$ c.m. 
frame; $x$ and $z$ are light-cone momentum fractions, with (see Ref. 
\cite{ourpaper} for exact relationships and further detail):
\be
x = \xbj  + {\cal O} \left(\frac{\kt^2}{Q^2} \right) \quad\quad
z = z_h   + {\cal O} \left(\frac{\kt^2}{Q^2} \right) \quad\quad
\bfp_\perp =  \bfP_T - z_h \, \bfk _\perp + 
{\cal O}\left(\frac{\kt^2}{Q^2} \right) \>. \label{appkin}
\ee
The elementary Mandelstam variables are given by
\bea
\hat s^2 &=& \frac{Q^4}{y^2} \left[ 1 - 4 \frac{\kt}{Q}\, \sqrt{1-y} \, 
\cos\varphi \right] + {\cal O} \left( \frac{\kt^2}{Q^2} \right) 
\label{eq:s2}\\
\hat u^2 &=& \frac{Q^4}{y^2} \, (1-y)^2 \left[ 1 - 4 \frac{\kt}{Q} \,
\frac{\cos\varphi}{\sqrt{1-y}} \right] +
{\cal O} \left( \frac{\kt^2}{Q^2} \right) \>,
\label{eq:u2}
\eea
where $\varphi$ is the azimuthal angle of the quark transverse momentum, 
$\bfk_\perp$ = $\kt(\cos\varphi,$ $\sin\varphi,0)$. Regarding angle 
definitions and notations we adopt throughout the paper the so-called 
``Trento conventions'' \cite{trento} (see also Fig. 3 of 
Ref. \cite{ourpaper}).  

The $\sin(\phi_h-\phi_S)$ weighted transverse single spin asymmetry,
measured by HERMES and COMPASS, which singles out the contribution of the 
Sivers function (\ref{rel}), is given by:
\bea
&& \!\!\!\!\!\!\!\! A^{\sin (\phi_h-\phi_S)}_{UT} = \label{hermesut} \\
&& \hskip-20pt \frac{\displaystyle  \sum_q \! \int \!\! 
{d\phi_S \, d\phi_h \, d^2 \bfk _\perp}\;
\Delta ^N \! f_{q/\pup} (x,\kt) \sin (\varphi -\phi_S) \; 
\frac{d \hat\sigma ^{\ell q\to \ell q}}{dQ^2} \;
J\; \frac{z}{z_h} \; D_q^h(z,p _\perp) \sin (\phi_h -\phi_S) }
{\displaystyle \sum_q \! \int \!\! {d\phi_S \, d\phi_h \, d^2 \bfk _\perp}\; 
f_{q/p}(x,\kt) \; \frac{d \hat\sigma ^{\ell q\to \ell q}}{dQ^2} \;
J\; \frac{z}{z_h} \; D_q^h(z,p _\perp) } \, \cdot \nonumber 
\eea
We shall use Eq. (\ref{hermesut}), in which we insert a parameterization for 
the Sivers functions, to fit the experimental data. 

We adopt the usual (and convenient) gaussian factorization for the 
distribution and fragmentation functions:
\be
f_{q/p}(x,k_\perp) = f_q(x) \, \frac{1}{\pi \langle\kt^2\rangle} \,
e^{-{\kt^2}/{\langle\kt^2\rangle}}
\label{partond}
\ee
and 
\be
D_q^h(z,p _\perp) = D_q^h(z) \, \frac{1}{\pi \langle p_\perp^2\rangle}
\, e^{-p_\perp^2/\langle p_\perp^2\rangle} \>,
\label{partonf}
\ee
with the values of $\langle k_\perp^2\rangle$ and $\langle p_\perp^2\rangle$ 
of Eq. (\ref{ktpar}). Isospin and charge--conjugation relations imply
\bea 
&& D_u^{\pi^+}(z) = D_d^{\pi^-}(z) =
   D_{\bar u}^{\pi^-}(z) = D_{\bar d}^{\pi^+}(z) \equiv D_{\rm fav}(z) 
\nonumber \\
&& D_u^{\pi^-}(z) = D_d^{\pi^+}(z) =
   D_{\bar u}^{\pi^+}(z) = D_{\bar d}^{\pi^-}(z) \equiv D_{\rm unfav}(z) 
\label{iso} \>.
\eea

The integrated parton distribution and fragmentation functions $f_q(x)$ and
$D_q^h(z)$ are taken from the literature, at the appropriate $Q^2$ values 
of the experimental data \cite{mrst01,kre}. 

We parameterize, for each light quark flavour $q=u,d,\bar u,\bar d$, the 
Sivers function in the following factorized form:
\be
\Delta^N \! f_ {q/\pup}(x,\kt) = 2 \, {\cal N}_q(x) \, h(\kt) \, 
f_ {q/p} (x,\kt)\; , \label{sivfac}
\ee
where
\bea
&&{\cal N}_q(x) =  N_q \, x^{a_q}(1-x)^{b_q} \,
\frac{(a_q+b_q)^{(a_q+b_q)}}{a_q^{a_q} b_q^{b_q}}\; , 
\label{siversx} \\
&&h(\kt) = \frac{2\kt \, M_0}{\kt^2+ M_0^2}\; \cdot
\label{siverskt}
\eea
$N_q$, $a_q$, $b_q$ and $M_0$ (GeV/$c$) are free parameters. 
$f_ {q/p} (x,\kt)$ is the unpolarized distribution function defined in 
Eq.~(\ref{partond}). Since $h(\kt) \le 1$ and since we allow the constant 
parameter $N_q$ to vary only inside the range $[-1,1]$ so that 
$|{\cal N}_q(x)| \le 1$ for any $x$, the positivity bound for the Sivers 
function is automatically fulfilled:
\be
\frac{|\Delta^N\!f_ {q/\pup}(x,\kt)|}{2 f_ {q/p} (x,\kt)}\le 1\; . \label{pos}
\ee

We have first attempted a fit of the HERMES and COMPASS data, taking into 
account 4 Sivers functions (for $u, d, \bar u$ and $\bar d$ quarks), for a
total of 13 parameters, like in Ref. \cite{ourpaper}. However, it turns out
that the available data are almost insensitive to the sea quark (and, in 
general, small $x$) contributions, which leads to largely undetermined 
parameters of the corresponding Sivers functions. Indeed, we have explicitely 
checked that various choices of $\Delta^N \! f_{\bar u/\pup}$ and 
$\Delta^N \! f_{\bar d/\pup}$ do not significantly affect the computation of 
$A_{UT}^{\sin(\phi_h-\phi_S)}$, Eq. (\ref{hermesut}), in the kinematical 
regions of the performed experiments. We have then neglected the 
contributions of these functions and considered only the contributions of 
$\Delta^N \! f_{u/\pup}$ and $\Delta^N \! f_{d/\pup}$, for a total of 7 free 
parameters:
\be
N_u \quad a_u \quad b_u \quad N_d \quad a_d \quad b_d \quad M_0.
\label{par}
\ee 

\begin{figure}[t]
\includegraphics[width=0.75\textwidth,bb= 20 30 570 450]{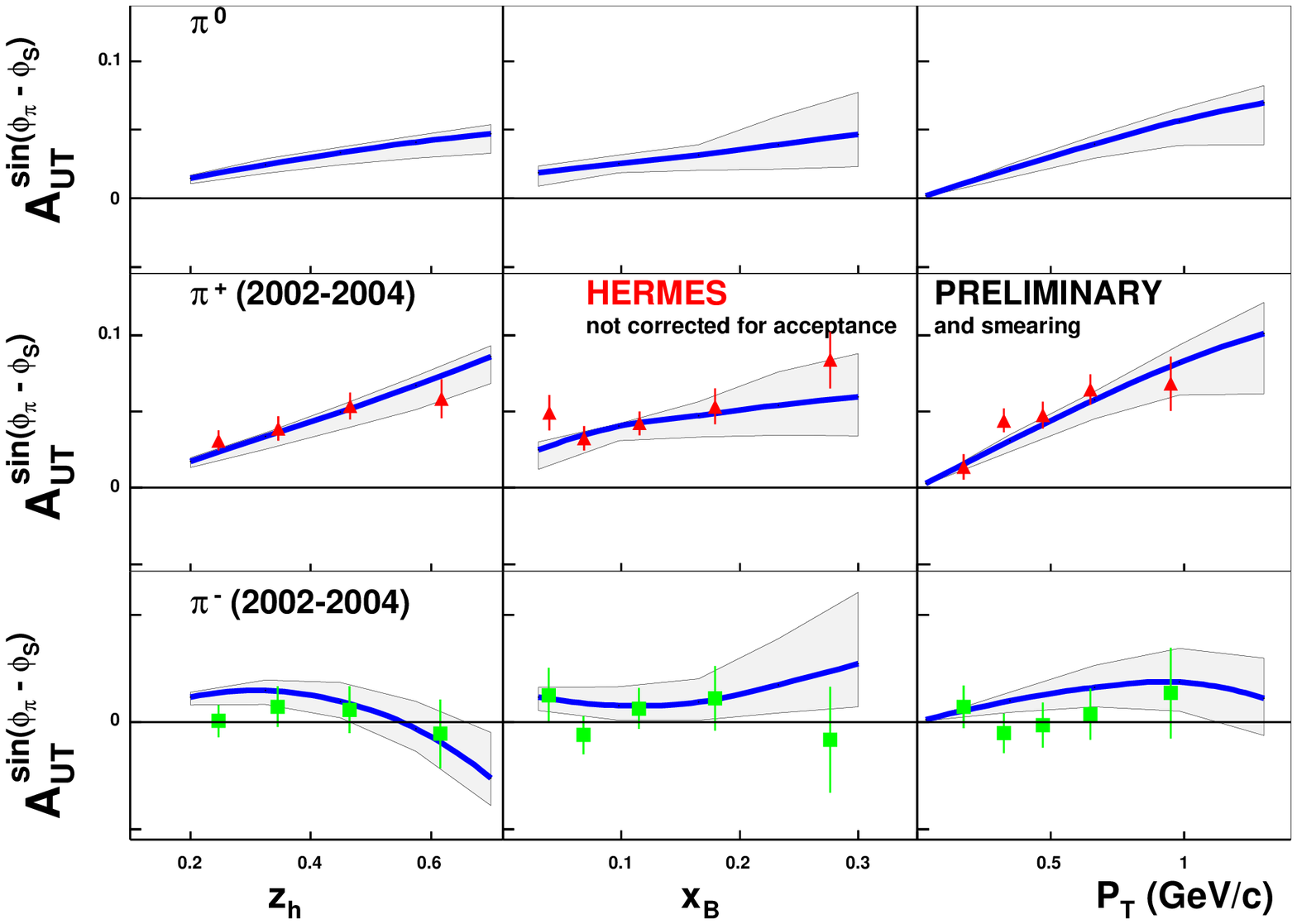}
\includegraphics[width=0.57\textwidth,angle=-90,bb= 20 30 570 450]{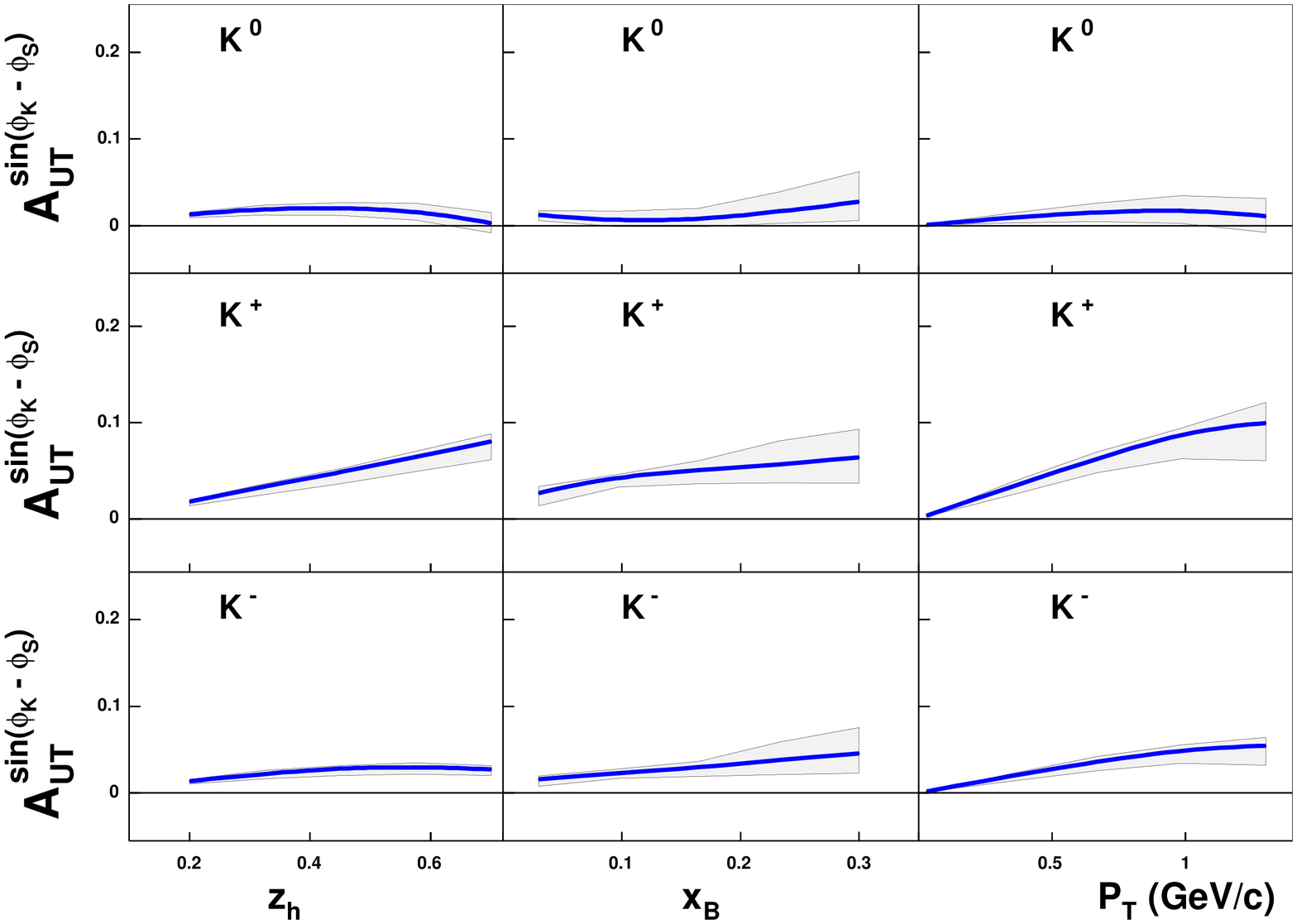}
\caption{\small
HERMES data on $A_{UT}^{\sin(\phi_\pi-\phi_S)}$ \cite{hermnew} for scattering
off a transversely polarized proton target and charged pion production. The 
curves are the results of our fit. The shaded area spans a region 
corresponding to one-sigma deviation at 90\% CL (see text for further detail). 
Predictions for $\pi^0$ (upper panel) and kaon (lower panels) asymmetries are 
also shown.}
\label{fig:authermes}
\end{figure}

\begin{figure}[b]
\includegraphics[width=0.8\textwidth,bb= 20 30 570 450]
{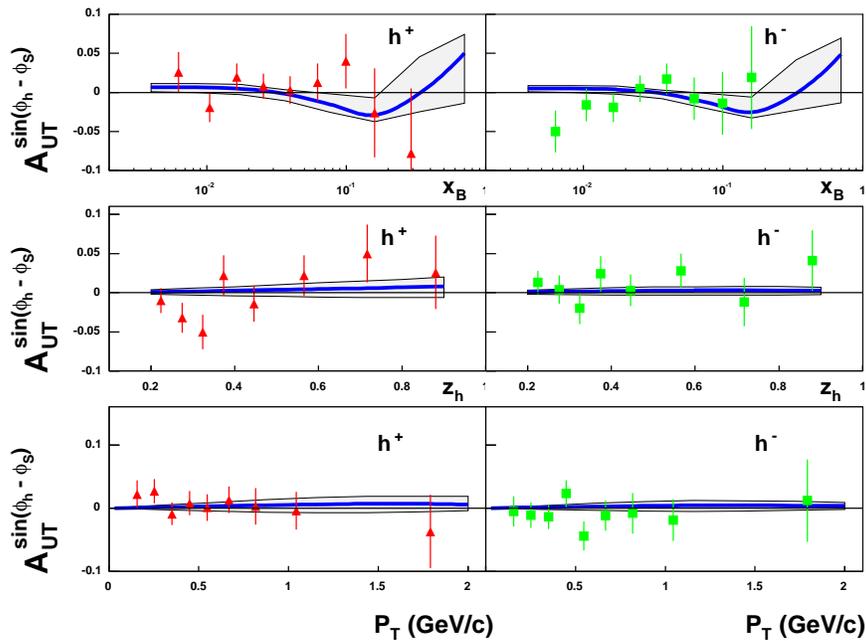}
\caption{\small
COMPASS data \cite{compnew} on $A_{UT}^{\sin(\phi_h-\phi_S)}$ for scattering 
off a transversely polarized deuteron target and the production of positively
($h^+$) and negatively ($h^-$) charged hadrons. The curves are
the results of our fit. The shaded area spans a region corresponding to 
one-sigma deviaton at 90\% CL (see text for further detail).}
\label{fig:autcompass}
\end{figure}

The results of our fits are shown in Figs. \ref{fig:authermes} and
\ref{fig:autcompass}. The weighted SSA $A_{UT}^{\sin(\phi_h-\phi_S)}$  
is plotted as a function of one variable at a time, either $z_h$ or $\xbj$ 
or $P_T$; an integration over the other variables has been performed 
consistently with the cuts of the corresponding experiment (see 
Ref. \cite{ourpaper} for further detail). The resulting best fit values 
of the parameters are reported in Table I. The shaded area in 
Figs. \ref{fig:authermes} and \ref{fig:autcompass} corresponds to one-sigma 
deviation at 90\% CL and was calculated using the errors (Table I) 
and the correlation matrix generated by MINUIT, minimizing and 
maximizing the function under consideration, in a 7-dimensional parameter 
space hyper-volume corresponding to one-sigma deviation.

In Fig. 1 we also show predictions, obtained using the extracted Sivers 
functions, for $\pi^0$ and $K$ production; data on these asymmetries might 
be available soon from HERMES collaboration.    

%
\begin{table}[t]
\begin{center}
\begin{tabular}{|ll|ll|}
\hline
$N_{u}$  = & $0.32  \pm  0.11$ & $N_{d}$ = & $-1.00  \pm  0.12$ \\
$a_{u}$  = & $0.29  \pm  0.35$ & $a_{d}$ = & $ 1.16  \pm  0.47$ \\
$b_{u}$  = & $0.53  \pm  3.58$ & $b_{d}$ = & $ 3.77  \pm  2.59$ \\
\hline
$M_0^2$ = & $0.32 \pm 0.25 \; {\rm (GeV}/c)^2$ & $\chi^2/{d.o.f.}$ = 
& $1.06$ \\
\hline
\end{tabular}
\end{center}
\caption{\small Best fit values of the parameters of the Sivers functions. 
Notice that the errors generated by MINUIT are strongly correlated, and should 
not be taken at face value; the significant fluctuations in our results are 
shown by the shaded areas in Figs. \ref{fig:authermes} and 
\ref{fig:autcompass}.    
\label{fitpar}}
\end{table}

\vspace{18pt}
\goodbreak
\nd
{\bf 3. Comparison with models and predictions for SIDIS processes}
\vskip 6pt

\begin{figure}[t]
\includegraphics[angle=-90,width=0.7\textwidth,bb= 30 40 340 450]
{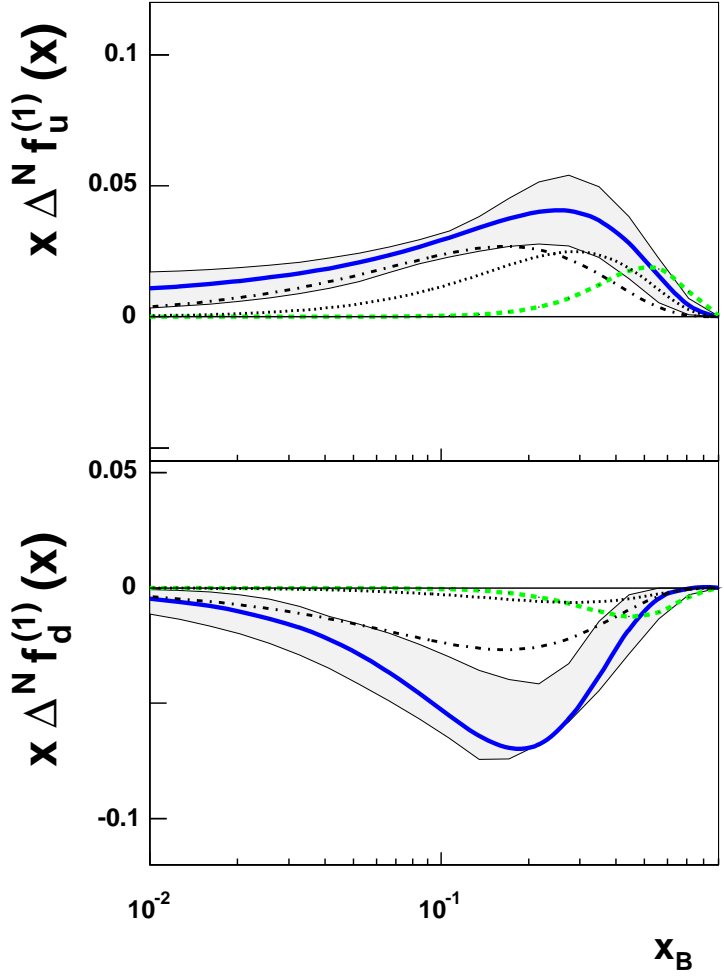}
\vskip 6cm
\caption{\small The $x$-dependence of the first $\bfk_\perp$ moment, 
according to Eq. (\ref{mom}), of our extracted Sivers functions. The solid 
line is obtained by using the central values of the parameters in Table I and 
the shaded area corresponds to varying the parameters within the shaded areas 
in Figs. 1 and 2. The dot-dashed, dashed and dotted lines show the first
$\bfk_\perp$ moments of the Sivers functions obtained respectively in  
Refs. \cite{efremov}, \cite{uf} and \cite{yuan}.}
\label{fig:sivers_functions.ps}
\end{figure}

The extracted Sivers functions for $u$ and $d$ quarks are shown in 
Fig. \ref{fig:sivers_functions.ps}, where we plot, for comparison with 
other results, the first $\bfk_\perp$ moment
\be
\Delta^N \! f_q^{(1)}(x) \equiv \int d^2 \, \bfk_\perp \, \frac{\kt}{4m_p} \, 
\Delta^N \! f_{q/\pup}(x, \kt) = - f_{1T}^{\perp (1) q}(x) \>. \label{mom}
\ee
The solid line corresponds to the central values in Table I and the shaded 
area corresponds to varying the parameters within the shaded areas  
in Figs. \ref{fig:authermes} and \ref{fig:autcompass}. The other curves 
show results from models or fits to different data \cite{efremov, uf, yuan}, 
as discussed below.

\begin{itemize}
\item
The $x$-dependences of both $\Delta^N\!f_{u/\pup}$ and $\Delta^N\!f_{d/\pup}$ 
-- as modeled in Eqs. (\ref{sivfac}-\ref{siverskt}) and shown in 
Fig. \ref{fig:sivers_functions.ps} -- appear to be rather well determined, 
keeping in mind that the data are essentially confined in the region 
$0.01 \lsim \xbj \lsim 0.2$. We notice that a 13-parameter fit -- including 
$\bar u$ and $\bar d$ contributions -- would lead to similar results; however, 
the values of $\Delta^N \! f_{\bar u/\pup}$ and $\Delta^N \! f_{\bar d/\pup}$, 
within their shaded areas, would be consistent with zero in the kinematical 
region of HERMES and COMPASS experiments. That is why we have not considered 
these contributions here.     
\item
The large-$x$ behaviour of the Sivers functions cannot be fixed by the 
existing data. According to the counting rules of Ref. \cite{bbs}, and 
keeping in mind that $\Delta^N \! f_{q/\pup}$ originates from the interference 
between {\it distribution amplitudes} with different proton helicities
\cite{old,bro}, one expects the large-$x$ behaviours
\be
\Delta^N\!f_{u/\pup} \sim \Delta^N\!f_{d/\pup} \sim (1-x)^4 \>. \label{count} 
\ee
JLab data will cover the appropriate region to help checking this prediction. 
\item
The dot--dashed line in Fig. \ref{fig:sivers_functions.ps} shows fit I of 
Ref. \cite{efremov}, where the $q_T/m_p$ weighted SIDIS asymmetries were 
fitted and the large $N_c$ relation \cite{drago, pob} was adopted:
\be
f_{1T}^{\perp u}(x, \kt) = - f_{1T}^{\perp d}(x, \kt) \>. \label{u-d}
\ee
Notice that their results are in qualitative agreement with ours and that
Eq. (\ref{u-d}) naturally turns out to be approximately true in our fit.  
\item
The dashed line in Fig. \ref{fig:sivers_functions.ps} plots the first 
$\bfk_\perp$ moment of the Sivers functions obtained in Ref. \cite{uf}, by 
fitting $A_N$ data in $\pup \, p \to \pi \, X$ processes; these data are 
mainly sensitive to large $x$ value, where -- again -- approximately opposite 
values of $\Delta^N\!f_{u/\pup}$ and $\Delta^N\!f_{d/\pup}$ seem to be 
favoured. However, as discussed in Ref. \cite{ourpaper}, the universality of 
the Sivers functions active in SIDIS and $\pup \, p \to \pi \, X$ processes 
is still an open issue. 
\item
Most theoretical models give a Sivers function for $u$ quarks much larger, 
in magnitude, than for $d$ quarks; this can be seen, for example, 
from the dotted curve in Fig. \ref{fig:sivers_functions.ps}, taken from
the computation, in the MIT bag model, of Ref. \cite{yuan}. The same is true 
for the Sivers functions obtained, within a spectator model with diquarks, 
in Refs. \cite{bsy}, \cite{lm} and \cite{ggo}. 
\end{itemize}

\vspace{12pt}
\goodbreak
\nd
{\it 3.1 $A_{UT}^{\sin(\phi_h-\phi_S)}$ at COMPASS with polarized hydrogen 
target}
\vskip 6pt

By inspection of Eq. (\ref{hermesut}) it is easy to understand our numerical 
results for the $u$ and $d$ Sivers functions. In fact one can see that for 
scattering off a hydrogen target (HERMES), one has 
\be
\left( A^{\sin (\phi_h-\phi_S)}_{UT}\right)_{\rm hydrogen} \sim
4 \, \Delta^N \! f_{u/\pup} \, D_u^h + \Delta^N \! f_{d/\pup} \, D_d^h \>,
\label{hydt}
\ee
while, for a scattering off a deuterium target (COMPASS),  
\be
\left( A^{\sin (\phi_h-\phi_S)}_{UT}\right)_{\rm deuterium} \sim
\left( \Delta^N \! f_{u/\pup} + \Delta^N \! f_{d/\pup} \right)
\left( 4 \, D_u^h + D_d^h \right) \>.
\ee
Opposite $u$ and $d$ Sivers contributions suppress COMPASS asymmetries for
any hadron $h$. These opposite contributions do not affect the $\pi^+$
asymmetry measured off a a hydrogen target, Eq. (\ref{hydt}): in this 
case the charge factor 4 and the favourite fragmentation function 
($D_u^{\pi^+} > D_d^{\pi^+}$) combine to make the first term of 
Eq. (\ref{hydt}) larger than the second one. The cancellation between the 
two terms is stronger for the $\pi^-$ asymmetry, because in this case 
the charge factor 4 in the first term of Eq. (\ref{hydt}) couples to the 
unfavoured fragmentation function ($D_u^{\pi^-} < D_d^{\pi^-}$). Similar 
arguments hold for the production of kaons and, in general, for the 
production of charged hadrons, which is dominated by pions.      

\begin{figure}[t]
\includegraphics[width=0.75\textwidth,bb= 20 30 570 450]
{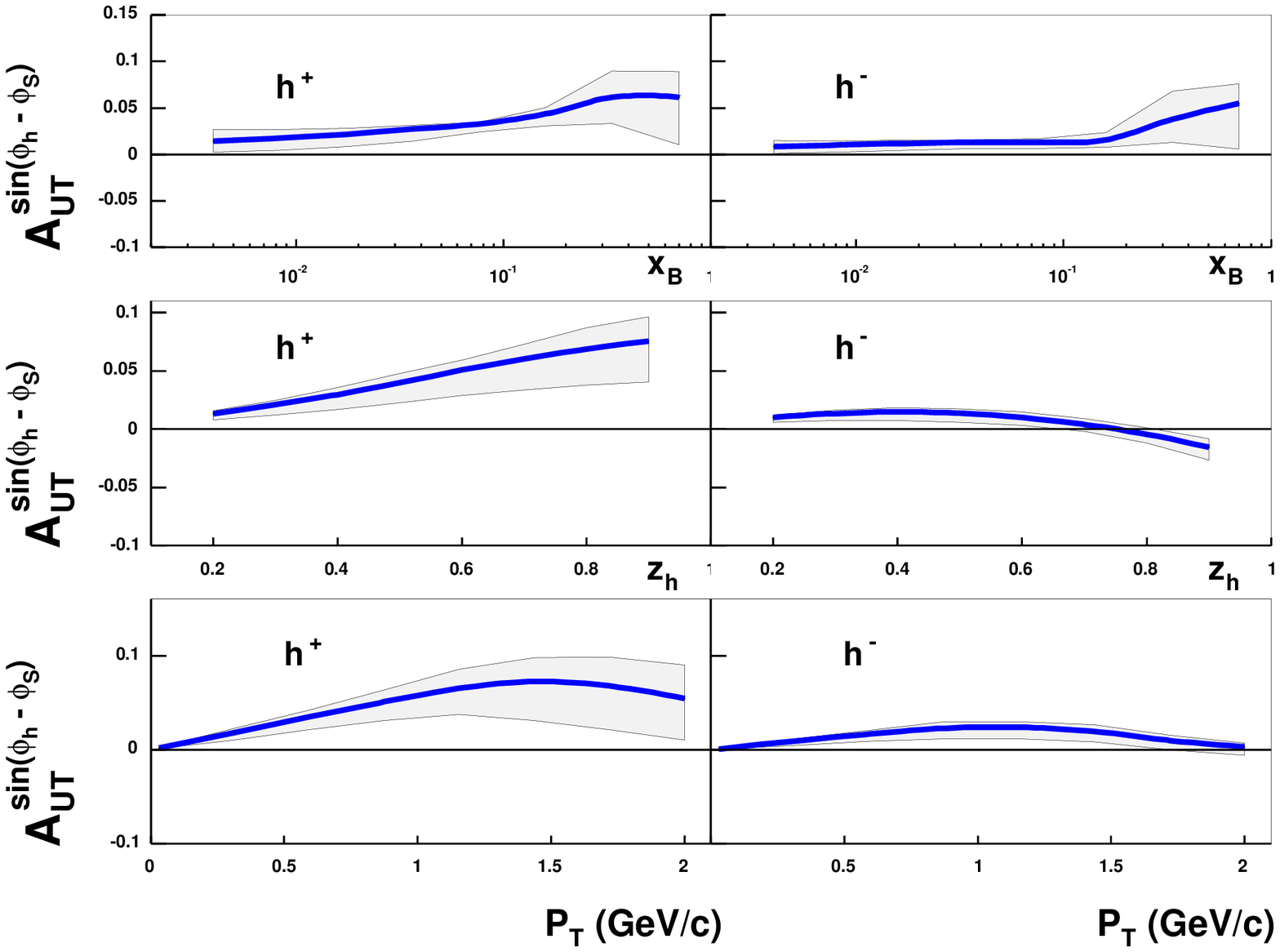}
\includegraphics[width=0.75\textwidth,bb= 20 30 570 450]
{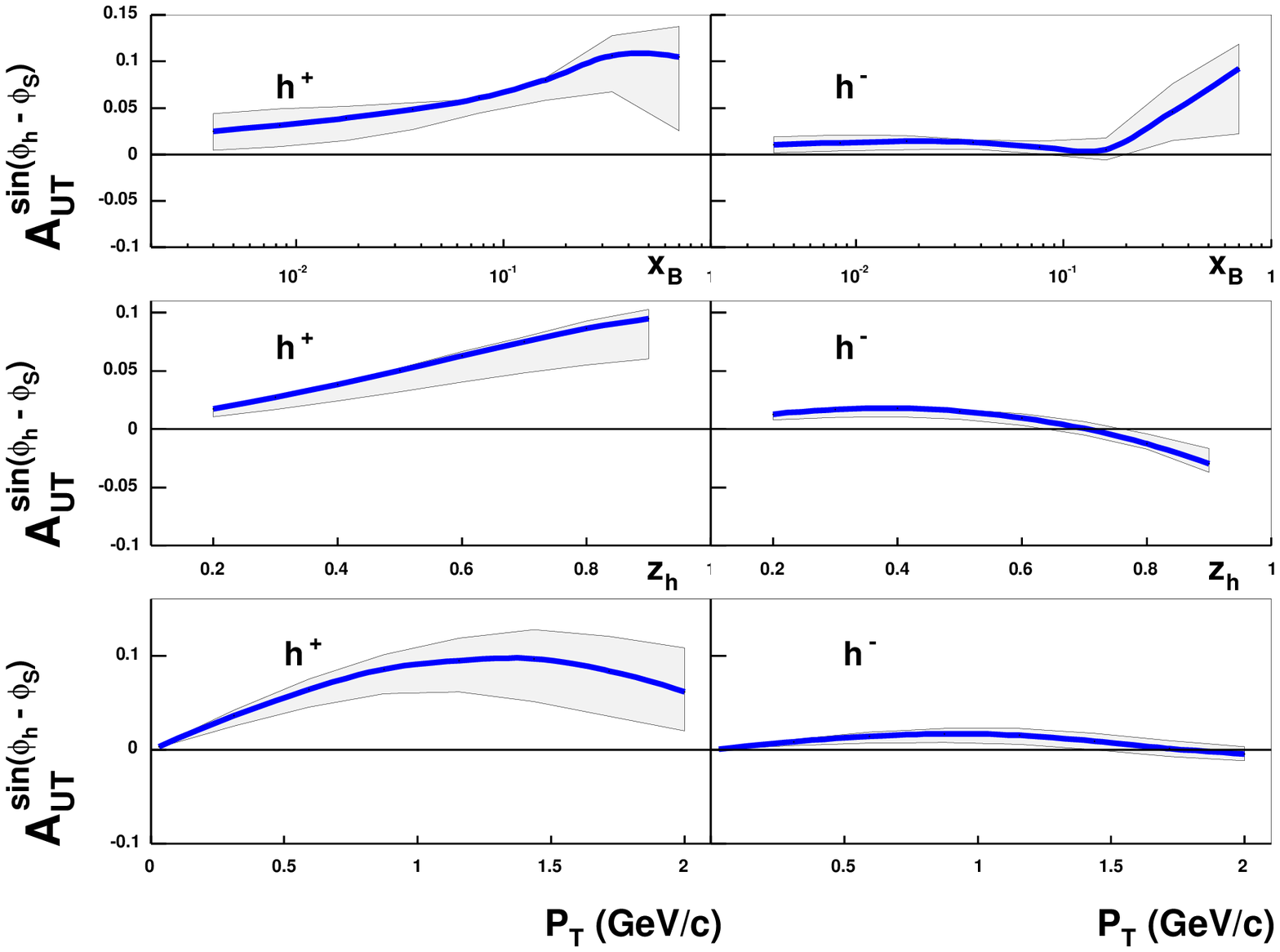}
\caption{\small 
Predictions for $A_{UT}^{\sin(\phi_h-\phi_S)}$ at COMPASS for scattering off 
a transversely polarized proton target and the production of positively
($h^+$) and negatively ($h^-$) charged hadrons. The plots in the upper panel
have been obtained by performing the integrations over the unobserved 
variables according to the standard COMPASS kinematical cuts \cite{ourpaper};
results with suggested new cuts, Eq. (\ref{new cuts}), are presented in the 
lower panel.}
\label{fig:autcompass_proton}
\end{figure}

However, the COMPASS collaboration will soon be taking data with a transversely
polarized hydrogen target. We can easily compute the expected results: 
adopting the same experimental cuts which were used for the deuterium target 
\cite{ourpaper} we obtain the predictions shown in the upper panel of 
Fig. \ref{fig:autcompass_proton}. The asymmetry is found to be around 5\%.
These expected values can be further increased by properly selecting the 
experimental data, thus excluding kinematical regions whose contribution to
the asymmetry is negligible. For example, selecting events with
\be
0.4 \le z_h \le 1 \quad\quad
0.2 \le P_T \le 1 \; {\rm GeV}/c \quad\quad
0.02 \le \xbj \le 1 \;,
\label{new cuts}
\ee 
yields the predictions shown in the lower panel of  
Fig. \ref{fig:autcompass_proton}. The asymmetry for positively charged hadrons
becomes larger, and, provided that enough statistics can be gathered, one 
expects a clear observation of a sizeable azimuthal asymmetry also for the 
COMPASS experiment.

\vspace{12pt}
\goodbreak
\nd
{\it 3.2 $A_{UT}^{\sin(\phi_h-\phi_S)}$ at JLab with polarized hydrogen 
target}
\vskip 6pt

Also JLab experiments are supposed to measure the SIDIS azimuthal asymmetry 
for the production of pions on a transversely polarised hydrogen target, at 
incident beam energies of 6 and 12 GeV. The kinematical region of this 
experiment is very interesting, as it will supply information on the 
behaviour of the Sivers functions in the large-$\xbj$ domain, up to 
$\xbj \simeq 0.6$. The experimental acceptance for JLab events 
at 6 GeV is constrained by \cite{priv}:
\bea
&& 0.4 \le z_h \le 0.7 \quad\quad
0.02 \le P_T \le 1 \; {\rm GeV}/c \quad\quad
0.1 \le \xbj \le 0.6 \label{JLab 6 GeV} \\
&& 0.4 \le y \le 0.85 \quad\quad
Q^2 \ge 1 \; ({\rm GeV}/c)^2 \quad\quad
W^2 \ge 4\; {\rm GeV}^2 \quad\quad
1 \le E_h \le 4\; {\rm GeV}. \nonumber
\eea
while, with an incident beam energy of 12 GeV, this becomes:
\bea
&& 0.4 \le z_h \le 0.7 \quad\quad
0.02 \le P_T \le 1.4 \; {\rm GeV}/c \quad\quad
0.05 \le \xbj \le 0.7 \label{JLab 12 GeV} \\
&& 0.2 \le y \le 0.85 \quad\quad
Q^2 \ge 1 \; ({\rm GeV}/c)^2 \quad\quad
W^2 \ge 4\; {\rm GeV}^2 \quad\quad
1 \le E_h \le 7\; {\rm GeV}. \nonumber
\eea
 
\begin{figure}[t]
\includegraphics[width=0.75\textwidth,bb= 20 30 570 450]
{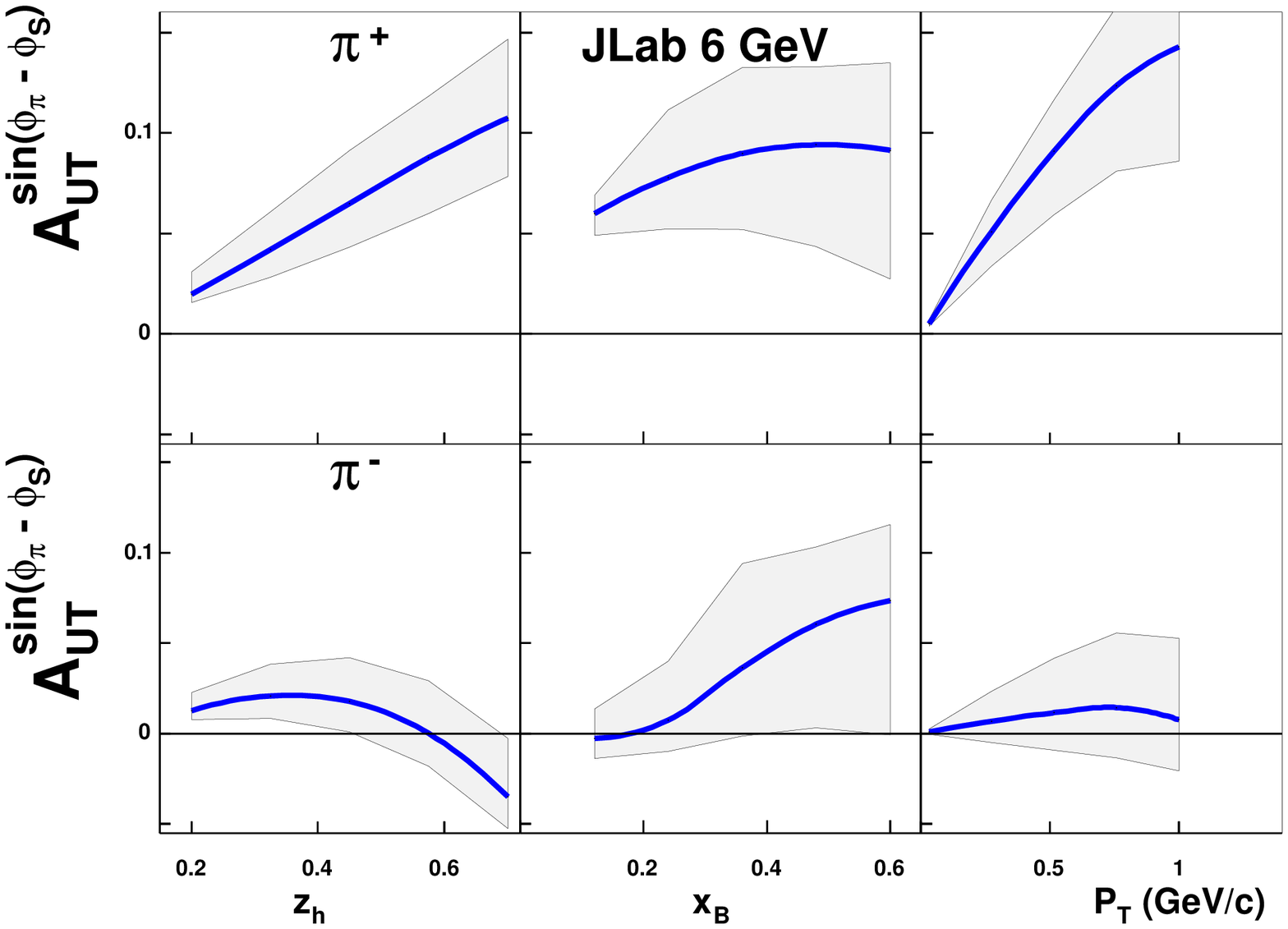}
\includegraphics[width=0.75\textwidth,bb= 20 30 570 450]
{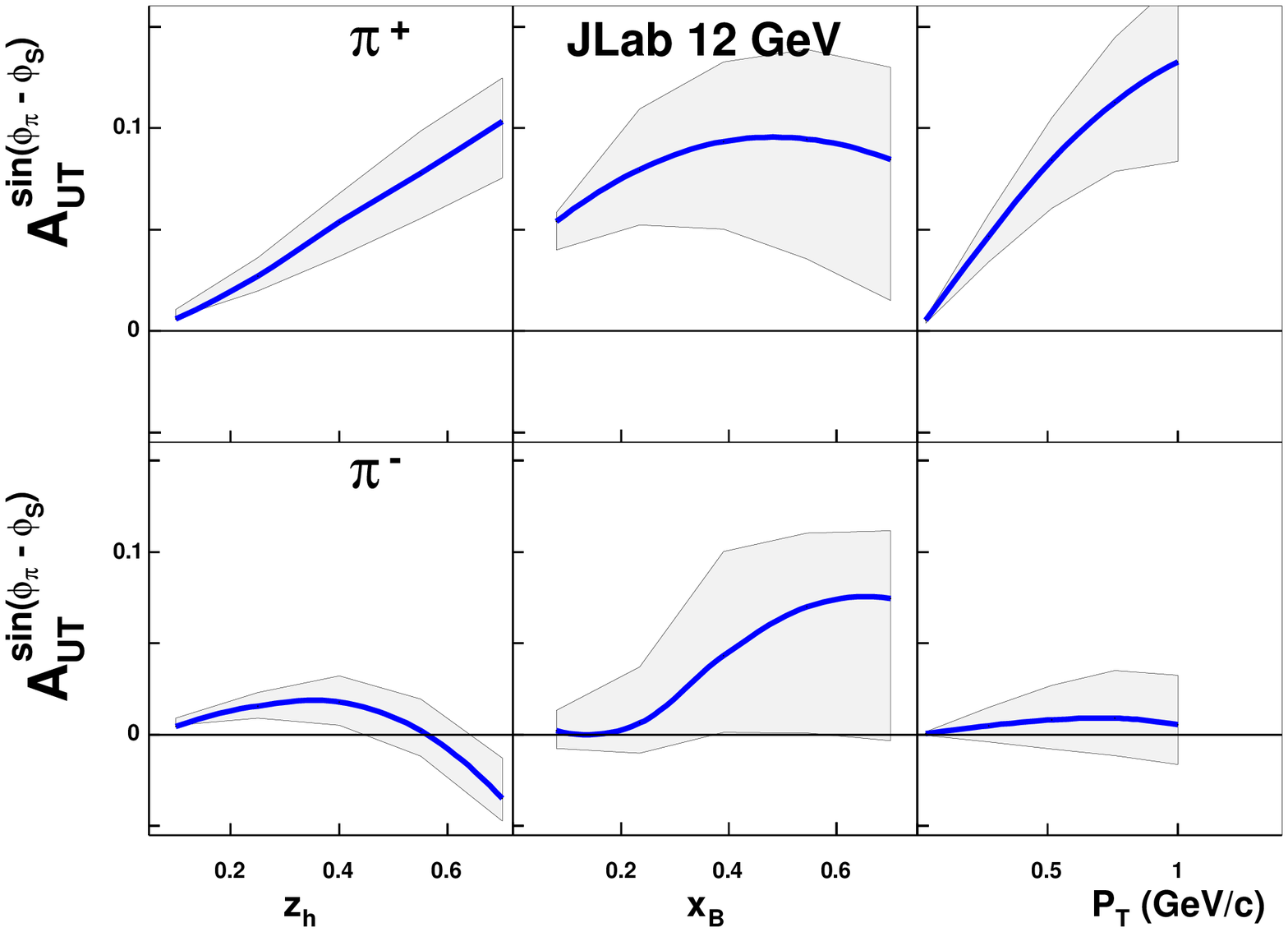}
\caption{\small
Predictions for $A_{UT}^{\sin(\phi_{\pi}-\phi_S)}$ at JLab for the production 
of $\pi^+$ and $\pi^-$ from scattering off a transversely polarized proton 
target. Integrations over the unobserved variables have been performed 
according to the kinematical ranges of Eqs. (\ref{JLab 6 GeV}) and 
(\ref{JLab 12 GeV}).}
\label{fig:autjlab}
\end{figure}

Imposing these experimental cuts we obtain the predictions shown in 
Fig. \ref{fig:autjlab}. A large and healthy azimuthal asimmetry for 
$\pi^+$ production should be observed. Similar results have been obtained 
also in an approach based on a Monte Carlo event generator \cite{aram}.  
However, one relevant comment is in order: 
\begin{itemize}
\item
As the region of high $\xbj$ is not covered by HERMES and COMPASS experiments, 
the predictions for the large-$\xbj$ dependence of the asymmetry are very 
sensitive to the few large-$\xbj$ data points of these two experiments. 
As a consequence, the results for JLab experiments may still change 
drastically in the region $0.4 \lsim \xbj \lsim 0.6$, and the asymmetry might 
be much smaller than presented in Fig. \ref{fig:autjlab}. This reflects in the 
wide shaded area at large $\xbj$ values. Conversely, the results
on $P_T$ and $z_h$ dependences are more stable as they depend on the  
$\xbj$-integrated Sivers function. Notice also the little dependence on the 
beam energy, consistent with the approximate factorized form of the numerator
and denominator of Eq. (\ref{hermesut}), which leads to cancellations in 
their ratio.

\end{itemize}

\vspace{18pt}
\goodbreak
\nd
{\bf 4. Transverse single spin asymmetries in Drell-Yan processes}
\vskip 6pt

Let us now consider the transverse single spin asymmetry, 
\be
A_N = \frac{d\sigma^\uparrow - d\sigma^\downarrow}
           {d\sigma^\uparrow + d\sigma^\downarrow} \> , \label{asy}
\ee
for Drell-Yan processes, $\pup \, p \to \ell^+ \, \ell^- \, X$, 
$\pup \, \bar p \to \ell^+ \, \ell^- \, X$ and 
$\bar p^{\,\uparrow} \, p \to \ell^+ \, \ell^- \, X$, where $d\sigma$ 
stands for
\be
\frac{d^4\sigma}{dy \, dM^2 \, d^2\bfq_T}
\ee
and $y$, $M^2$ and $\bfq_T$ are respectively the rapidity, the squared 
invariant mass and the transverse momentum of the lepton pair in the initial
nucleon c.m. system. The cross section can eventually be integrated over 
some of these variables, according to the kinematical configurations of the 
experiments.    

In such a case the single spin asymmetry (\ref{asy}) can only originate 
from the Sivers function and is given (selecting the region with 
$q_T^2 \ll M^2, \> q_T \simeq k_\perp$) by \cite{noi2} 
\be
A_N = \frac
{\sum_q e_q^2 \int d^2\bfk_{\perp q} \, d^2\bfk_{\perp \bar q} \>
\delta^2(\bfk_{\perp q} + \bfk_{\perp \bar q} - \bfq_T) \>
\Delta^N \! f_{q/\pup}(x_q, \bfk_{\perp q}) \>
f_{\bar q/p}(x_{\bar q}, \bfk_{\perp \bar q})}
{2 \sum_q e_q^2 \int d^2\bfk_{\perp q} \, d^2\bfk_{\perp \bar q} \>
\delta^2(\bfk_{\perp q} + \bfk_{\perp \bar q} - \bfq_T) \>
f_{q/p}(x_q, \bfk_{\perp q}) \>
f_{\bar q/p}(x_{\bar q}, \bfk_{\perp \bar q})} \>, \label{ann}
\ee
where $q = u, \bar u, d, \bar d, s, \bar s$ and
\be
x_q = \frac{M}{\sqrt s} \, e^y \quad\quad\quad
x_{\bar q} = \frac{M}{\sqrt s} \, e^{-y} \>. 
\ee
Eq. (\ref{ann}) explicitely refers to $\pup\,p$ processes, with obvious 
modifications for $\pup \, \bar p$ and $\bar p^{\, \uparrow} \, p$ ones.  

Inserting into Eq. (\ref{ann}) the Sivers functions extracted from our fit 
to SIDIS data and {\it reversed in sign} according to Eq. (\ref{DIS-DY}), 
we obtain the predictions shown in Figs. 6 and 7. Fig. 6 shows the 
value of $A_N$ as a function of $M$ and $x_F = x_q - x_{\bar q}$,
for RHIC configurations: the lepton pair transverse momentum $\bfq_T$ 
has been integrated in the range $0 \leq q_T \leq 1$ GeV/$c$, while the 
rapidity variable $y$ and the lepton pair invariant mass $M$ have been 
integrated according to the experimental situations, as indicated in the 
legenda. The integration over the azimuthal angle of $\bfq_T$ has been 
performed, to avoid cancellations, as in Ref. \cite{noi2}, that is 
integrating over $\phi_{q_{_T}}$ in the range $[0,\pi/2]$ only, or, 
alternatively, taking into account the change of sign in the different 
production quadrants. In either case, the $\phi_{q_{_T}}$ integration gives
an overall factor $2/\pi$. As the shaded areas in previous figures, the 
closed areas correspond to the uncertainty in our determination of the Sivers 
functions.  

Fig. 7 shows the same plots for the PAX experiment \cite{pax} planned at 
the proposed asymmetric $p \, \bar p$ collider at GSI: $\bfq_T$ has been
integrated over the same range as for RHIC predictions, while $y$ and $M$
as indicated in the legenda. Results for $\pup \, \bar p$ and 
$\bar p^{\, \uparrow} \, p$ processes are identical, due to charge 
conjugation invariance. Notice that in our configuration the polarized 
proton or antiproton always moves along the $(+\hat{\bfz})$--direction.       

The correct interpretation of these results requires some further 
considerations.
\begin{itemize}
\item
In our computations we have used the value of
$\langle\kt^2\rangle = 0.25 \;({\rm GeV}/c)^2$, obtained from an analysis 
of SIDIS data \cite{ourpaper}; such a value is certainly appropriate for
consistently computing spin asymmetries in SIDIS processes, in the 
$\gamma^* p$ c.m. frame, as we have done in Section 3.1 and 3.2. 
This value naturally corresponds to the intrinsic motion of partons confined  
in a nucleon, simply according to uncertainty principle, and describes well 
the $\bfP_T$ dependences of measured cross sections, up to $P_T \simeq 1$
GeV/$c$. In addition, as we have seen, it allows an understanding of the 
azimuthal asymmetries, which would otherwise vanish.     
\item
However, when considering other processes, as the inclusive production of 
hadrons or leptons in $p\,p$ or $p \, \bar p$ interactions, we know that 
higher order QCD corrections, like the threshold resummation of large 
logarithms due to soft gluon emission \cite{werner}, lead to large 
$K$-factor enhancements of the cross sections. Our $\bfk_\perp$ unintegrated 
approach to the description of hard scattering processes within a 
generalization of the QCD factorization theorem \cite{uf,noi3}, can be 
considered as an effective model which not only takes into account the 
original partonic intrinsic motion (related to parton confinement), but also, 
to some extent, the intrinsic $\bfk_\perp$ built via soft gluon emission.
Indeed, the values of $\langle k_\perp^2 \rangle$ used in Ref. \cite{uf} 
in order to describe the data on the unpolarized $p \, p \to \pi \, X$ 
processes are higher than the values used here, and those requested 
for the Drell-Yan cross-section might be even higher. The average 
$\langle\kt^2\rangle$ estimate of $0.25 \;({\rm GeV}/c)^2$ might be at most 
adequate to explain the Drell-Yan cross section up to $q_T \lsim$ 1 GeV/$c$, 
but would badly fail above that value. 
\item
For the reasons explained above, consistently with our approach expected to 
hold in the $\kt \simeq P_T \simeq q_T$ region, in our predictions for $A_N$, 
Figs. 6 and 7, we have integrated over $\bfq_T$ up to $q_T$ = 1 GeV/$c$.
In addition, we notice that the value of $A_N$, as given by Eq. (\ref{ann}), 
is little sensitive to the chosen value of $\langle\kt^2\rangle$: while both 
the numerator and the denominator of Eq. (\ref{ann}) greatly vary with  
$\langle\kt^2\rangle$, their ratio does much less so. We then consider our 
predictions for $A_N$, assuming the validity of the relation (\ref{DIS-DY}), 
safe and significant. 
\end{itemize}
       
\vspace{18pt}
\goodbreak
\nd
{\bf 5. Comments and conclusions}
\vskip 6pt

We have considered the most recent data from polarized SIDIS processes which 
single out the Sivers effect, namely the $A_{UT}^{\sin(\phi_h-\phi_S)}$ 
transverse single spin azimuthal asymmetry, measured by HERMES \cite{hermnew} 
and COMPASS \cite{compnew} collaborations for charged hadron production. 
Assuming a Gaussian factorization of the $\kt$ and $p_\perp$ dependence of 
all distribution and fragmentation functions, together with a most simple 
parameterization of the $x$-dependence of the unknown Sivers functions, we 
have exploited the data to extract information on 
$\Delta^N \! f_ {u/\pup}(x,k_\perp)$ and $\Delta^N \! f_ {d/\pup}(x,k_\perp)$. 

For the first time, the amount and quality of the experimental results 
allow a significant, although still limited in $x$-range, estimate of the 
Sivers functions for $u$ and $d$ quarks; these turn out to be definitely 
different from zero, well inside the positivity bound of Eq. (\ref{pos}) 
and almost opposite to each other. This last feature, predicted theoretically 
in some models \cite{drago, pob}, explains naturally and is related to the 
small asymmetry observed by COMPASS in scatterings of muons off a deuteron 
target. 

According to the general strategy of combining new experimental information 
with the computation and prediction of new expected results, the extracted 
functions have been used to compute $A_{UT}^{\sin(\phi_h-\phi_S)}$ in 
other experiments. It turns out that, contrary to the results obtained 
off a polarized deuteron target,
a sizeable $h^+$ asymmetry should be measured by COMPASS 
collaboration once they switch, as planned, to a transversely polarized 
hydrogen target; a careful choice of the kinematical region of the selected
events would help in further increasing the numerical value of the asymmetry
for positively charged hadron production.  

Large values of $A_{UT}^{\sin(\phi_h-\phi_S)}$ are expected at JLab, both 
in the 6 and 12 GeV operational modes, for $\pi^+$ inclusive production; 
in particular the $z_h$ and $P_T$ dependence of the asymmetry seems to be
stable and reliable, while the $\xbj$ dependence shows large uncertainties 
due to the lack of HERMES and COMPASS information in this region.
JLab experiments have the unique features of exploring the large $\xbj$ 
behaviour of the quark distribution functions, where predictions from 
QCD counting rules, Eq. (\ref{count}), could be tested.  
  
The Sivers effect was believed for some time to be forbidden by QCD 
time reversal properties \cite{colv}; however, this proved to be incorrect
\cite{col} after an explicit model showed the existence of a non zero
Sivers function \cite{bro}. The original proof of the vanishing of the 
Sivers effect turned into the relation of Eq. (\ref{DIS-DY}), which predicts 
opposite values for the Sivers functions measured in SIDIS and Drell-Yan 
processes. We have then used this basic QCD relation and computed the single 
spin asymmetries in Drell-Yan processes given in Eq. (\ref{ann}); these can 
only be generated by the Sivers functions, since no fragmentation functions 
are needed to describe this process. We have used the same functions 
as extracted from SIDIS data, with opposite signs. The predicted $A_N$ could 
be measured at RHIC in $p\,p$ collisions and, in the long range, at the 
proposed PAX experiment at GSI \cite{pax}, in $p\,\bar p$ interactions. 
It would provide a clear and stringent test of basic QCD properties.       

A phenomenological study of the Sivers asymmetry -- the correlation between
the intrinsic $\bfk_\perp$ of partons and the proton spin -- is now possible, 
thanks to the existing experimental information and more which will soon be 
available. Basic properties of the QCD proton structure can and will be 
clarified. A good control of the Sivers mechanism will help in learning and 
understanding about other fundamental partonic spin properties, like the 
transversity distribution \cite{bdr} and the Collins mechanism \cite{colv}.
    
\vspace{18pt}
\goodbreak
\nd
{\bf Acknowledgements}
\vskip 6pt
We would like to thank Elke Aschenauer, Harut Avakian and Delia Hasch for 
fruitful discussions. We acknowledge the support of the European 
Community--Research Infrasctructure Activity under the FP6 
``Structuring the European Research Area'' programme (HadronPhysics, contract 
number RII3-CT-2004-506078). U.D. and F.M. acknowledge partial support 
by MIUR (Ministero dell'Istruzione, dell'Universit\`a e della Ricerca) under 
Cofinanziamento PRIN 2003.





\begin{figure}[t]
\includegraphics[width=0.45\textwidth,angle=90,bb= 540 50 200 770]
{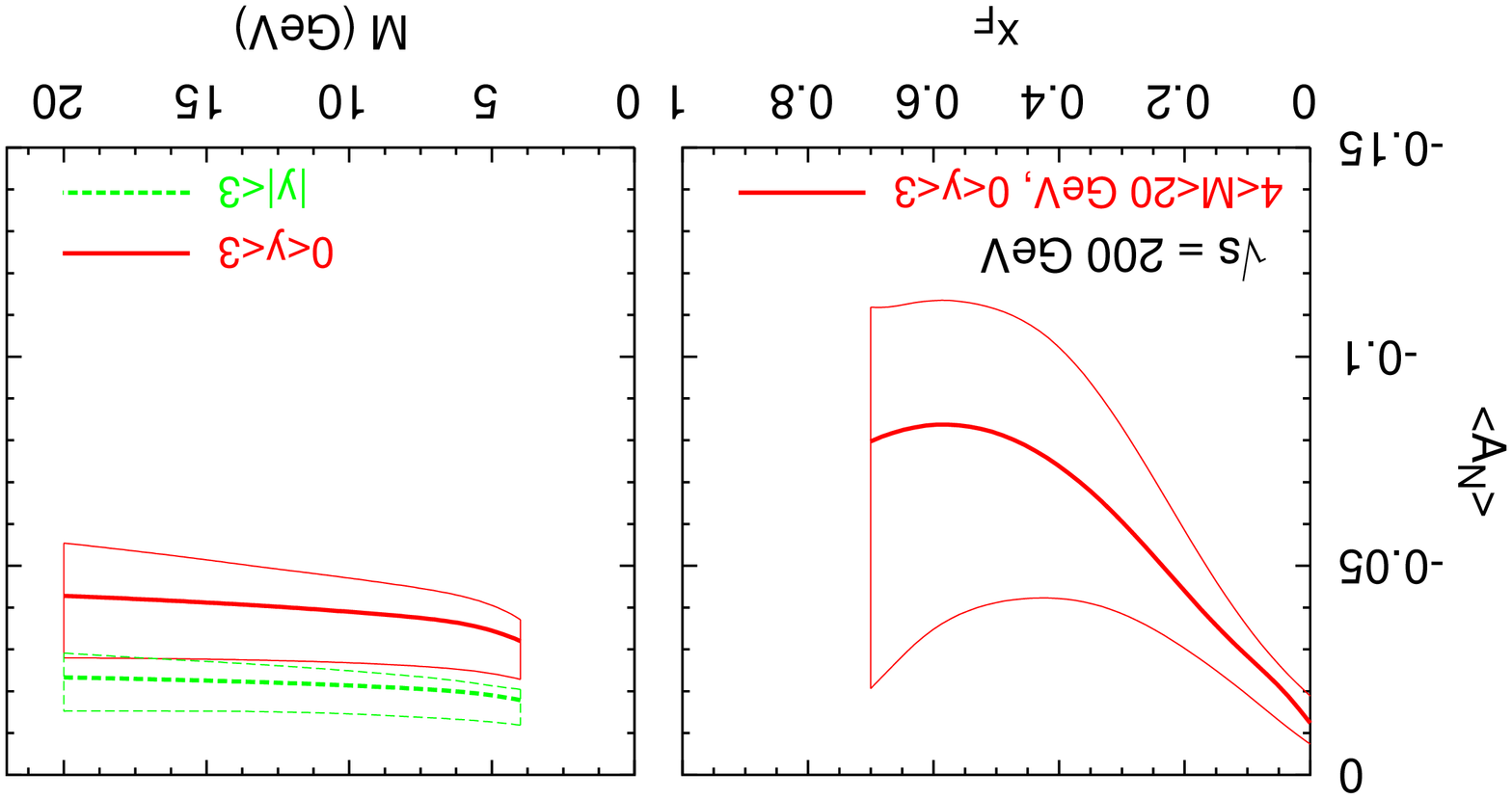}
\caption{\small
Predictions for single spin asymmetries in Drell-Yan processes at RHIC,
$\pup \, p \to \ell^+ \, \ell^- \, X$, according to Eq. (\ref{ann}) of the 
text. The lepton pair transverse momentum has been integrated in the range
$0 \leq q_T \leq 1$ GeV/$c$; $A_N$ is plotted as a function of $x_F$ (left)
and $M$ (right), with integration over the other variable as indicated 
in the legenda (see text for further details). Values of $\langle A_N \rangle$
for negative $x_F$ are negligible, at RHIC kinematics.}
\label{fig:anrhic}
\end{figure}

\begin{figure}[t]
\includegraphics[width=0.45\textwidth,angle=90,bb= 540 50 200 770]
{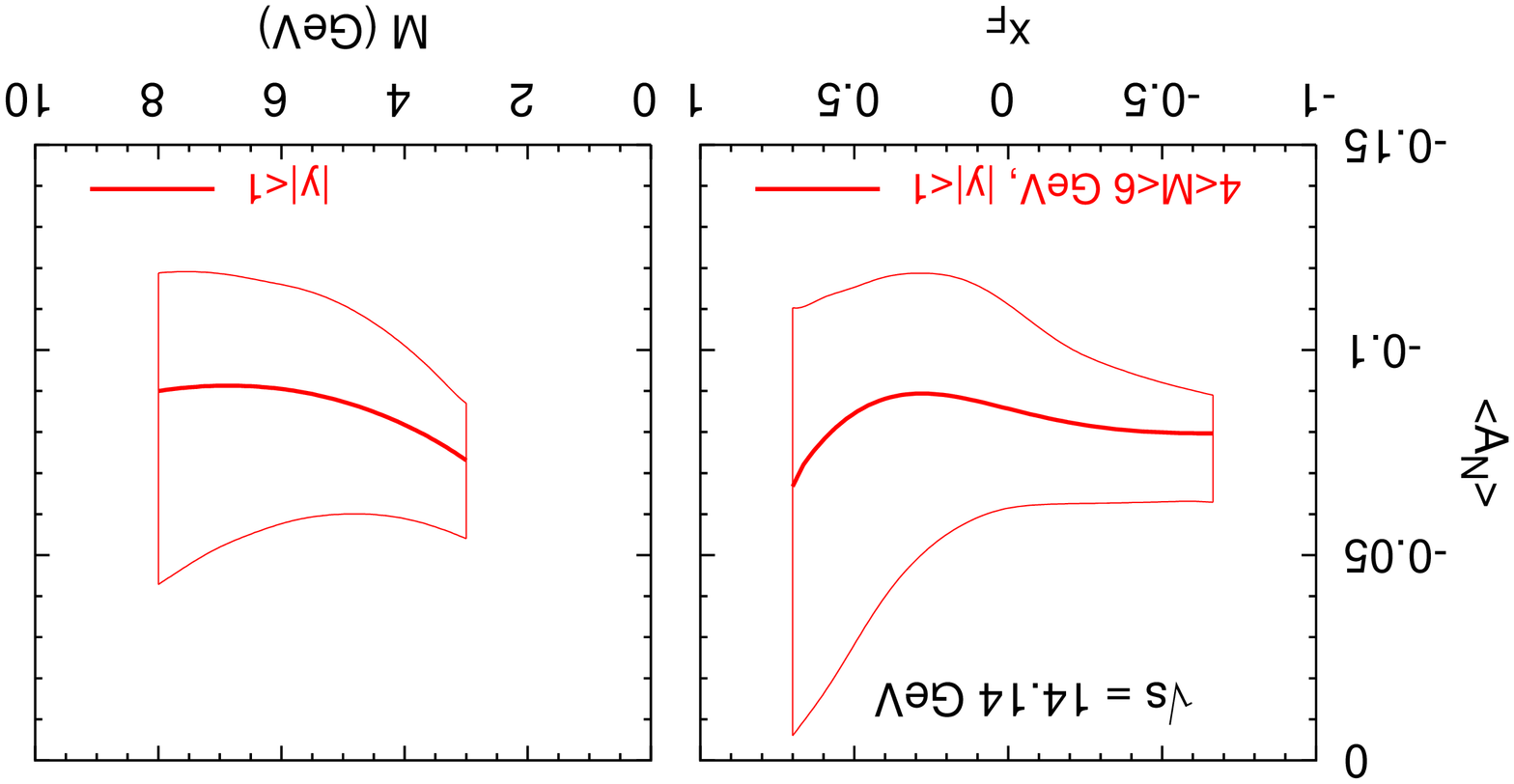}
\caption{\small
Predictions for single spin asymmetries in Drell-Yan processes at GSI,
$\pup \, \bar p \to \ell^+ \, \ell^- \, X$, according to 
Eq. (\ref{ann}) of the text. The lepton pair transverse momentum has been 
integrated in the range $0 \leq q_T \leq 1$ GeV/$c$; $A_N$ is plotted as a 
function of $x_F$ (left) and $M$ (right), with integration over the other 
variable as indicated in the legenda (see text for further details). The 
results for the $\bar p^{\uparrow} \, p  \to \ell^+ \, \ell^- \, X$ process 
are the same.}
\label{fig:anpax}
\end{figure}


\vskip 36pt
\baselineskip=6pt
\begin{thebibliography}{99}
\small
\vspace{6pt}
\setlength{\parskip}{6pt}
\bibitem {ourpaper}
\vskip-8pt
  M. Anselmino, M. Boglione, U. D'Alesio, A. Kotzinian, F. Murgia and
  A. Prokudin, \PR {D71} (2005) 074006
\bibitem{cahn} 
\vskip-8pt
  R.N. Cahn, \PL{B78} (1978) 269; \PR{D40} (1989) 3107
\bibitem{trento}
\vskip-8pt
  A. Bacchetta, U. D'Alesio, M. Diehl and C.A. Miller, \PR {D70} (2004) 117504
\bibitem{hermUT}
\vskip-8pt
  HERMES Collaboration, A. Airapetian {\it et al.}, \PRL {94} (2005) 012002; 
  U.~Elschenbroich, G.~Schnell and R.~Seidl (on behalf
  of HERMES Collaboration), e-Print Archive: hep-ex/0405017;  
  talk by N. Makins (on behalf of HERMES Collaboration),
  Transversity Workshop, Athens, Greece, October 6-7, 2003
\bibitem{siv}
\vskip-8pt
  D. Sivers, \PR{D41} (1990) 83; {\bf D43} (1991) 261
\bibitem{compUT} 
\vskip-8pt
  COMPASS Collaboration, P. Pagano, talk delivered at the SPIN2004 Symposium, 
  Trieste, Italy, October 10-16, 2004, e-Print Archive: hep-ex/0501035\\
  COMPASS Collaboration, R. Webb, talk delivered at BARYONS04, Oct 25-29 2004, 
  Palaiseau, France, e-Print Archive: hep-ex/0501031
\bibitem{hermnew}
\vskip-8pt
  HERMES Collaboration, M. Diefenthaler, talk delivered at DIS 2005, Madison,
  Wisconsin (USA), April 27 -- May 1, e-Print Archive: hep-ex/0507013 
\bibitem{compnew}
\vskip-8pt
  COMPASS Collaboration, V.Yu. Alexakhin {\it et al.}, \PRL {94} (2005) 202002 
\bibitem{col}
\vskip-8pt
  J.C. Collins, \PL {B536} (2002) 43
\bibitem{noi2}
\vskip-8pt
  M. Anselmino, U. D'Alesio and F. Murgia, \PR {D67} (2003) 074010
\bibitem{mrst01} 
\vskip-8pt 
  A.D. Martin, R.G. Roberts, W.J. Stirling and R.S. Thorne, 
  \PL {B531} (2002) 216
\bibitem{kre}
\vskip-8pt
  S. Kretzer, \PR {D62} (2000) 054001 
\bibitem{efremov}
\vskip-8pt
  A.V. Efremov, K. Goeke, S. Menzel, A. Metz and P. Schweitzer,
  \PL {B612} (2005) 233
\bibitem{uf}
\vskip-8pt
  U. D'Alesio and F. Murgia, \PR {D70} (2004) 074009 
\bibitem{yuan}
\vskip-8pt
  F. Yuan, \PL {B575} (2003) 45
\bibitem{bbs}
\vskip-8pt
  S.J. Brodsky, M. Burkardt and I. Schmidt, \NP {B441} (1995) 197
\bibitem{old}
\vskip-8pt
  M. Anselmino, M. Boglione and F. Murgia, \PL {B362} (1995) 164
\vskip-8pt
\bibitem{bro} 
\vskip-8pt
  S.J. Brodsky, D.S. Hwang and I. Schmidt, \PL {B530} (2002) 99
\bibitem{drago}
\vskip-8pt
  M. Anselmino, V. Barone, A. Drago and F. Murgia,
  e-Print Archive: hep-ph/0209073; A. Drago, \PR {D71} (2005) 057501
\bibitem{pob}
\vskip-8pt
  P.V. Pobilytsa, e-Print Archive: hep-ph/0301236 
\bibitem{bsy}
\vskip-8pt
  A. Bacchetta, A. Sch\"afer and J-J. Yang, \PL {B578} (2004) 109 
\bibitem{lm}
\vskip-8pt
  Z. Lu and B.Q. Ma, \NP {A741} (2004) 200
\bibitem{ggo}
\vskip-8pt
  L.P. Gamberg, G.R. Goldstein and K.A. Oganessyan, \PR {D67} (2003) 071504 
\bibitem{priv}
\vskip-8pt
  H. Avakian, private communication
\bibitem{aram}
\vskip-8pt
  A. Kotzinian, e-Print Archive: hep-ph/0504081
\bibitem{pax}
\vskip-8pt
  PAX Collaboration, e-Print Archive: hep-ex/0505054
\bibitem{werner}
\vskip-8pt
  D. de Florian and W. Vogelsang, \PR {D71} (2005) 114004
\bibitem{noi3}
\vskip-8pt
  M. Anselmino, M. Boglione, U. D'Alesio E. Leader and F. Murgia, 
  \PR {D71} (2005) 014002 
 \bibitem{colv}
\vskip-8pt
  J.C. Collins, \NP {B396} (1993) 161
\bibitem{bdr}
\vskip-8pt
  For a comprehensive recent review paper on transversity, see, {\it e.g.} 
  V. Barone, A. Drago and P. Ratcliffe, {\it Phys. Rept.} {\bf 359} (2002) 1
\end{thebibliography}
\end{document}